# Композитная модель анализа данных сейсмического мониторинга при ведении горных работ на примере Кукисвумчоррского месторождения АО "Апатит"

*ФГБОУ ВО «Санкт-Петербургский горный университет» (Санкт-Петербург, Россия)*

*Аннотация:* Геомеханический мониторинг состояния массива горных пород – это активно развивающийся раздел геомеханики, в котором на данный момент практически невозможно выделить единую методологию и подходы для решения задач, сбора и анализа данных при разработке систем сейсмического мониторинга состояния массива. При ведении горных работ изменениям подвергаются все природные факторы. В процессе отработки массива горных пород наиболее явно проявляются изменения состояния структурных неоднородностей: раскрываются имеющиеся естественные структурные неоднородности; происходят подвижки по разрывным нарушениям (разломам); образуются новые, техногенные нарушения (трещины), которые сопровождаются изменением естественного напряженного состояния различных блоков массива. Разработанный метод оценки результатов мониторинга геомеханических процессов в массиве горных пород на примере Объединенного Кировского рудника КФ АО «Апатит» позволил решить одну из главных задач системы геомониторинга – осуществлен прогноз зон возможного возникновения опасных проявлений горного давления.

*Ключевые слова:* кластеризация, геомониторинг, сейсмика, анализ данных, машинное обучение, композитные модели

## Введение

Проводимые исследования являются дальнейшим развитием идеи, предложенной авторами статьи [1], об «использовании методов машинного обучения при работе с математическими моделями временных рядов, базирующихся на данных сейсмического мониторинга».

В задачу геомеханического мониторинга входит поиск взаимосвязей различного рода между широким перечнем природных и технических факторов, определяющих состояние техногенно-нарушенного массива горных пород. Можно выделить две основные группы таких факторов. Так, например, в работе А.А. Козырева с соавторами [7] этим факторам даются следующие определения: «Природные факторы – это сумма свойств пород, слагающих массив, структурных неоднородностей массива и естественного природного поля напряжений. Технические факторы – это совокупность методов ведения горных работ, порядок строительства объектов, применяемая система разработки, характеристики горных выработок и др.».

Одним из способов оценки вероятности проявления опасных геодинамических процессов является использование математических моделей, основанных на данных сейсмического мониторинга. Так как массив горных пород является сложной динамической системой, для построения таких моделей целесообразно использовать как пространственные координаты, так и компоненты временного ряда сейсмической активности. Последовательность разрушений, которые появляются в процесс эксплуатации месторождения, можно представить в виде некоторого временного ряда, состоящего из дискретных событий. Каждое из этих событий задается координатой во времени и характеристикой, описывающей степень разрушения массива, например, величиной энергии сигнала [13]. В данной статье в качестве такого временного ряда предлагается использовать сейсмическую последовательность, зарегистрированную с помощью системы контроля состояния массива горных пород. Для локализации пространственной зоны подготовки очага разрушения используются методы кластеризации, сопоставляющие каждому дискретному событию свой пространственный кластер.

Объектом настоящего исследования является Кукисвумчоррское месторождение, разрабатываемое Объединенным Кировским рудником КФ АО «Апатит». По данным П.А. Корчака и С.А. Жуковой «Мониторинг сейсмичности на подземных рудниках осуществляется с помощью автоматизированной системы контроля состояния массива (АСКСМ). За время наблюдения за сейсмичностью в массиве горных пород были выявлены закономерности:
• реакция массива на массовые взрывы;
• сезонный рост сейсмичности в периоды таяния снега;
• рост сейсмичности перед обрушением консоли налегающих пород.
Технологические процессы при ведении горных работ оказывают существенное влияние на сейсмический режим рудников. К настоящему времени на рудниках ОАО «Апатит» документально зарегистрировано около 40 горных ударов» [8].

**Методология**

Как прогнозируется уже в ближайшем будущем ожидается ряд технологических прорывов, отражающих глубинные технологические изменения, которые приведут к трансформации традиционного промышленного производства [11, 12, 28]. При этом необходимо учитывать, что увеличение объемов добычи полезных ископаемых в сложных горно-геологических условиях тесно связано с анализом основных параметров напряженно-деформированного состояния (НДС) массива горных пород.

Достоверный анализ и моделирование НДС массива горных пород базируется на применении различных математических моделей [2, 6]. Так, с точки зрения теории, предложенной в рамках иерархической модели, критерием формирования очага разрушения является нарушение условий стационарности/квазистационарности моделируемого процесса (например, Пуассона) [4]. Однако, отметим, что при данном подходе учитываются только статистические параметры, и игнорируются, например, физико-механические параметры исследуемой среды. Это негативно влияет на качество работы используемой прогностической модели и не позволяет учитывать особенности моделируемого месторождения.

Таким образом, получение качественных и достоверных результатов при реализации таких моделей без непосредственного участия эксперта предметной области исследований, всегда затруднено, так как он выбирает соответствующие стратегии предварительной обработки данных, тип используемой модели, ее параметры и набор критериев [3]. Предлагаемый в данной работе алгоритм генерирует, с помощью математической модели, управляемой данными сейсмического мониторинга, или так называемый data-driven model [35], собственный набор критериев формирования очага разрушения. Полученный набор учитывает как особенности ведения горных работ на месторождении, так и физико-механические свойства массива горных пород.

Искомую математическую модель можно разработать, используя как одиночную модель машинного обучения, так и гибридный (композитный) подход. В настоящее время идентификация data-driven моделей со сложной гетерогенной структурой остается нерешенной пока проблемой. Структура композитной модели может быть представлена в виде ориентированного ациклического графа (DAG), а наиболее подходящий вариант структуры может быть разработан с использованием оптимизационных подходов.

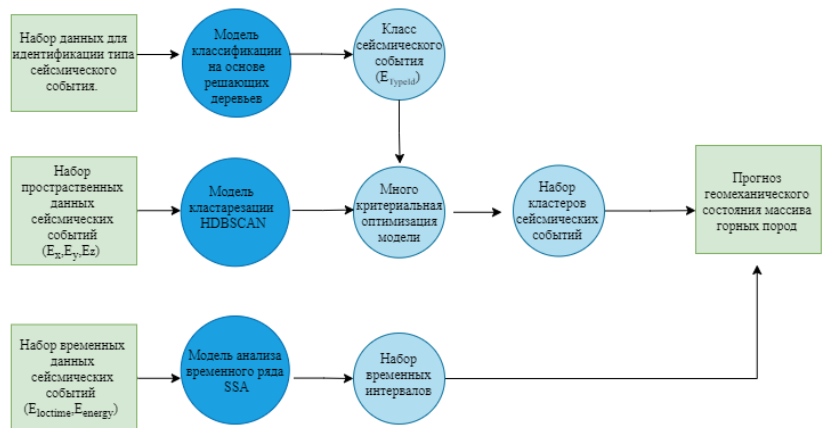

Рисунок 1 – Композитная модель прогноза опасных геодинамических явлений
*(Составлена авторами)*

Элементами этого графа являются модели машинного обучения (предназначенные для классификации или дополнительные модели для побочных задач, например, кластеризации).

Таким образом, задачей исследования является поиск кластеров – "очагов разрушения горной породы". Для такой задачи в области пространства-времени-энергии сейсмических событий, на основе проведенных исследований был разработан алгоритм системы анализа и прогноза геомеханического состояния техногенно-нарушенного горного массива (рис. 1). Последнийзаключается в одновременном применении различных моделей машинного обучения в единой композитной модели, наиболее подходящих для определенного типа данных, отражающих изменчивость наблюдаемой системы в каждой из компонент пространства. Оценка качества итоговой модели в работе определяется введением критериев, необходимых для организации процесса многокритериальной оптимизации.

1. **Критерий Silhouette**. Силуэтом выборки называется средняя величина силуэта объектов данной выборки; она показывает насколько среднее расстояние до объектов своего кластера отличается от среднего расстояния до объектов других кластеров. Эта величина принадлежит диапазону [-1,1]. Значения, близкие к -1, соответствуют плохим (разрозненным) кластеризациям; значения, близкие к нулю, означают, что кластеры пересекаются и накладываются друг на друга; значения, близкие к 1, соответствуют "плотным", четко выделенным кластерам, то есть, чем больше силуэт выборки, тем более четко выделены кластеры, представляющие собой компактные, плотно сгруппированные облака точек.

2. **Доли ложных (FAR) и пропущенных тревог (MAR)**. Важнейшим элементом решения проблемы детектирования аномалий является определение временного окна детектирования. В рамках предложенных критериев качества, предсказанные аномалии внутри окна детектирования воспринимаются только как один истинный положительный результат. Отсутствие прогнозируемых аномалий внутри окна обнаружения воспринимается только как один ложный отрицательный результат. Предсказанные точки за пределами окон обнаружения определяются как ложные срабатывания. В данном исследовании приняты 3 значения ширины окна детектирования:

- краткосрочный горизонт прогнозирования - 6 часов до и после наступления события;
- среднесрочный горизонт прогнозирования - 48 часов до и после наступления события;
- долгосрочный горизонт прогнозирования - 168 часов до и после наступления события.

Далее рассмотрим два из трех методов, используемых в итоговой композитной модели: Метод Singular Spectrum analysis (SSA) для анализа временных рядов сейсмического мониторинга, предназначенный для моделирования поведения в системы в пространстве времени-энергии, и метод кластеризации HDBSCAN, предназначенный для поиска кластеров-"очагов разрушения горной породы", который отвечает за моделирование пространственной изменчивости системы. Базовым вариантом метода SSA при анализе временных рядов является декомпозиция исходного ряда на простые компоненты, которые могут быть аппроксимированы с помощью периодических функций или полиномов низкой степени [24-27]. Полученное разложение может служить основой прогнозирования как самого временного ряда, так и его отдельных составляющих. Для анализа временного ряда сейсмического мониторинга выбирается параметр L, отвечающий за «ширину» окна. Выбор значения параметра L зависит от исследователя и его предметных знаний о системе, порождающей выбранный временной ряд. Затем, на основе ряда, строится матрица Ганкеля, где в качестве её столбцов используются векторы длины L. Задачей выбранного вида разложения является представление исходного ряда в виде суммы компонент. Такой метод рекомендуется применять для выделения тренда, циклических компонент и построения на основе выбранных компонент некоторой аппроксимации исходного временного ряда.

Анализ данных сейсмического мониторинга и кластеризация сейсмических событий требуют применения современных методов математического моделирования. В процессе изучения сейсмического режима Кукисвумчоррского месторождения за длительный период было установлено, что кластеры сейсмических событий и механизм их образования имеют схожие черты с другими месторождениями, в частности они: «приурочены к местам активного ведения горных работ, к дизъюнктивным нарушениям в консоли пород висячего бока, а также могут образовываться под влиянием других факторов как природных, так и техногенных» [5].

Классический подход использования методов кластеризации – это подход, «основанный на представлениях о физических процессах, происходящих в процессе нагружения и последующей деформации горного массива, для решения задачи идентификации очаговых зон наиболее целесообразным кажется использование иерархического или графового подхода. Иными словами, данные методы являются методами "строгой" кластеризации» [17,29]. При этом в ряде исследований утверждается, что использование методов "нестрогой" кластеризации более предпочтительно для моделирования случайных процессов [9, 14,16,29].

Окончательно, в работе был применен алгоритм HDBSCAN (Hierarchical Density-Based Spatial Clustering of Applications with Noise) [12,18,19]. Данный алгоритм является иерархическим пространственным алгоритмом кластеризации данных с шумом на основе использования плотности распределения [15,20,21].

**Результаты и обсуждения**

В качестве данных для валидации работы предложенного алгоритма используются даты зафиксированных горных ударов за период 2009 – 2018 гг (табл. 1). Отметим, что с целью соблюдения конфиденциальности, пространственные координаты были изменены.

Таблица 1. **Зафиксированные горные удары (по данным карточек горных ударов).**

| Координата сейсмического события по оси $Ox$ | Координата сейсмического события по оси $Oy$ | Координата сейсмического события по оси $Oz$ | Время наступления зафиксированного геодинамического события. |
|---|---|---|---|
| 0,456 | 0,413 | 0,325 | 13.05.2009 |
| 0,381 | 0,464 | 0,228 | 21.10.2010 |
| 0.603 | 0.441 | 0.293 | 27.01.2016 |

*(Составлена автором)*

Длина анализируемого временного ряда составляет 365 дней, ширина временного окна была выбрана равной 30 дням (параметр L), а в качестве наблюдаемой величины используется среднее значение (за один день) всех сейсмических событий со значением энергии в диапазоне от $10^2$ до $10^5$ Дж. Данный диапазон был обоснован анализом распределения сейсмических событий по величине энергии с января 2009 года по декабрь 2020 года, а также на основании экспертной оценки сотрудников Научного центра Геомеханики и проблем горного производства Санкт-Петербургского горного университета. В качестве оси OX выбрана временная шкала сейсмического мониторинга, а в качестве оси OY шкала величины среднего значения сейсмических событий (Дж).

Параметр предполагаемой периодичности процесса формирования горного удара был выбран равным 15 дням. Так как его значение составляет 50% от длины окна, то это означает наличие двух подпроцессов в рамках одного месяца. И если в первом подпроцессе (первые 15 дней) не было опасных проявлений горного давления (ГД), то вероятность их появления во втором подпроцессе (вторые 15 дней) изменяется в соответствии с ростом экспоненциальной функции. Восстановленные по первой элементарной матрице временные ряды для 2009, 2010 годов приведены на рис. 2, а для 2016 года — на рис. 3. Первая элементарная матрица отвечает за тренд сейсмической активности. Как видно из данных рис. 2, 13 апреля 2009 года и 25 августа 2010 года достигаются минимальные значения тренда сейсмической активности в выбранные периоды наблюдений.

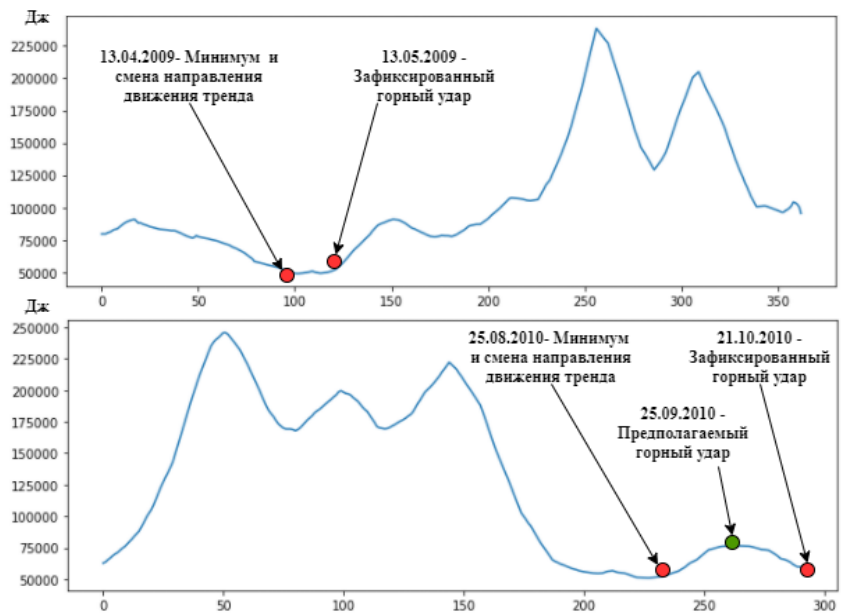

Рисунок 2 – Тренд временного ряда сейсмического мониторинга за 2009 и 2010 год соответственно *(Составлены авторами)*

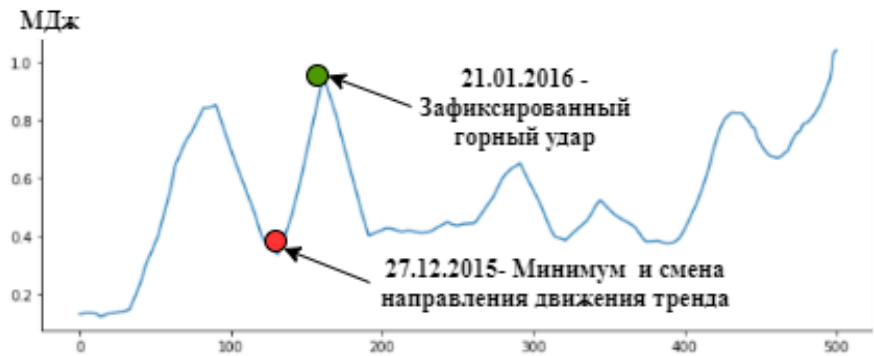

Рисунок 3 – Тренд временного ряда сейсмического мониторинга за 2016 год
*(Составлен авторами)*

На рис. 2 приведена «сумма» первого и второго случаев (2009 и 2010 год). С одной стороны, имеется повтор ситуации 2009 года (падение до минимального значения тренда сейсмической активности) и точка смены направления роста тренда 27 декабря 2015 года (рис. 3). Следовательно, анализ тренда сейсмической активности, полученного с помощью метода SSA, позволяет явным образом установить связь между опасным проявлениями ГД, в виде горного удара от 21.01.2016, и сейсмической активностью, выраженной нисходящим трендом.

***Таким образом, опасному проявлению ГД может предшествовать плавное снижение тренда до точки глобального минимума тренда с последующим изменением в направлении тренда. В соответствии с предположением о цикличности процесса, для выбранных параметров (ширина окна и периодичность процесса) время до наступления опасного проявления ГД события равно 15 и 30 дням соответственно.***

**Использование композитной модели для анализа данных сейсмического мониторинга за 2020 год**

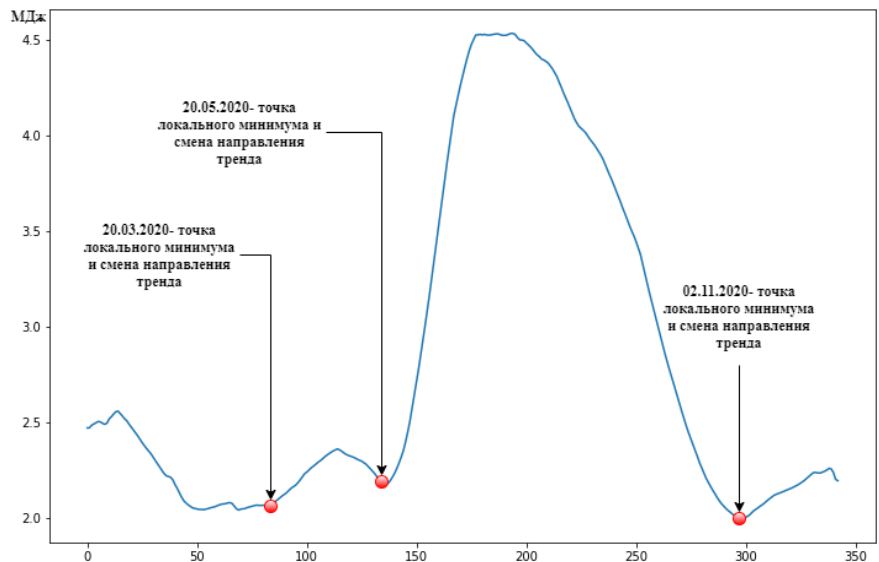

Рисунок 4 – Тренд временного ряда сейсмического мониторинга за 2020 год
*(Составлн авторами)*

На рис. 4, можно выделить 3 ключевые точки в тренде:
• 20 марта 2020 года достигается минимальное значение тренда сейсмической активности на текущий момент наблюдений. Однако, вероятность наступления горного удара – умеренная, так как росту тренда предшествует период стационарности (отсутствие роста среднего значения тренда), равный длине окна (30 дней). Данный период можно интерпретировать как стабилизацию процессов деформации в массиве горных пород, что позволяет нейтрализовать возможные опасные проявления ГД.
• 20 мая 2020 года достигается локальный минимум тренда сейсмической активности, и начинается резкий рост тренда в противоположном направлении. Вероятность наступления горного удара в следующие 30 дней достаточно велика, так как процесс достиг своего локального минимума, и начался рост сейсмической активности.
• 2 ноября 2020 года достигается глобальный минимум значения тренда сейсмической активности за весь период наблюдений, а затем начинается умеренный рост тренда сейсмической активности. Вероятность наступления горного удара в следующие 30 дней крайне велика, так как процесс достиг своего минимального исторического значения, и начался рост тренда сейсмической активности.

Таблица 2. **Сводная таблица прогнозов времени наступления потенциальных горных ударов**

| Время появления точки локального минимум и смены тренда | Время потенциального горного удара. Степень риска-умеренная | Время потенциального горного удара. Степень риска-крайне высокая | Время наступления зафиксированного геодинамического события |
|---|---|---|---|
| 20.03.2020 | 06.04.2020 | 21.04.2020 | 08.04.2020 |
| 20.05.2020 | 06.06.2020 | 20.06.2020 | 06.07.2020 |
| 02.11.2020 | 17.11.2020 | 03.12.2020 | 04.12.2020 |

*(Составлена авторами)*

В двух из трех зафиксированных горных ударах удалось спрогнозировать приблизительное (с разницей в 1-2 дня) время наступления горного удара (табл.2). Однако, для случая произошедшего 06.07.2020 ошибка прогноза составила 15 дней, что требует уточнения прогноза. Далее был рассмотрен детально случай и восстановлен исходный временной ряд, с использованием данных с 01.01.2020 по 05.07.2020 (за один день до фактического горного удара).

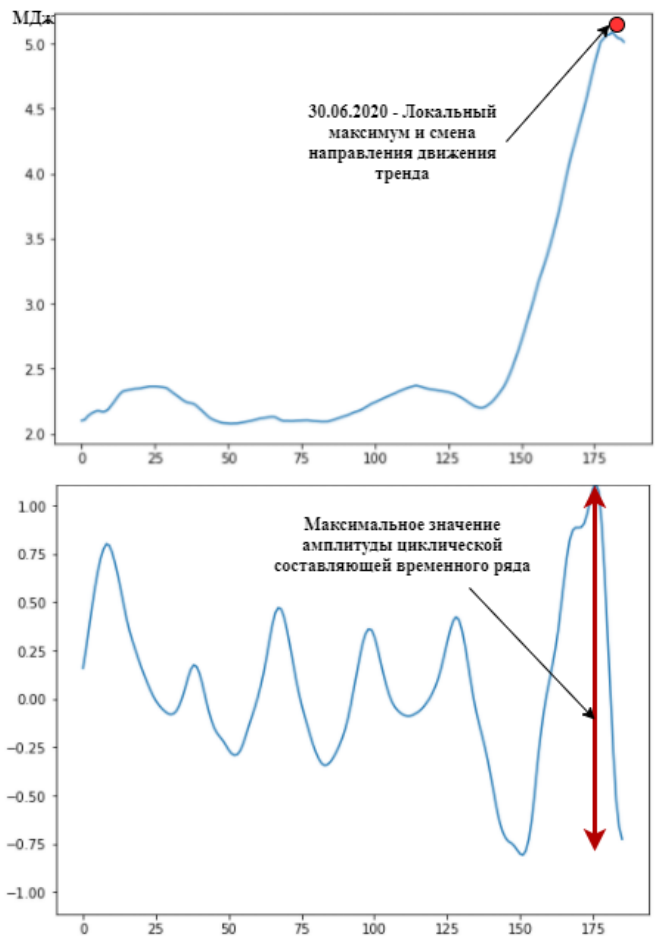

Рисунок 5 – Тренд и циклическая компонента временного ряда сейсмического мониторинга за период с 01.01.2020-05.07.2020 *(Составлен авторами)*

Как видно из рис. 5 в точке 30.06.2020 достигается локальный максимум, и начинается смена направления движения тренда. Так же в период с 30.06.2020 по 05.07.2020 зафиксировано максимальное значение амплитуды циклической составляющей. Совокупность этих факторов позволяет спрогнозировать наступление потенциального горного удара 06.07.2020. Таким образом, можно сделать вывод, что анализ тренда целесообразно дополнять анализом циклической составляющей.

**Анализ результатов моделирования кластеров сейсмических событий на основе данных сейсмического мониторинга за 2020 год**

В ряде работ отмечено: «Существенная часть подходов к прогнозным оценкам проявлений опасного горного давления основывается на следующей концепции: по мере разрушения горной породы происходит формирование нескольких стадий разрушения, с постепенным переходом от одной стадии к другой» [30, 31]. Размеры и диапазон трещин, которые образуются в результате деформации и разрушения массива горных пород, могут варьироваться от миллиметра до десятых долей метра. В реальности, наиболее часто, опасные проявления ГД выражаются в виде горных ударов и других проявлений техногенной сейсмичности (рис. 6 и 7).

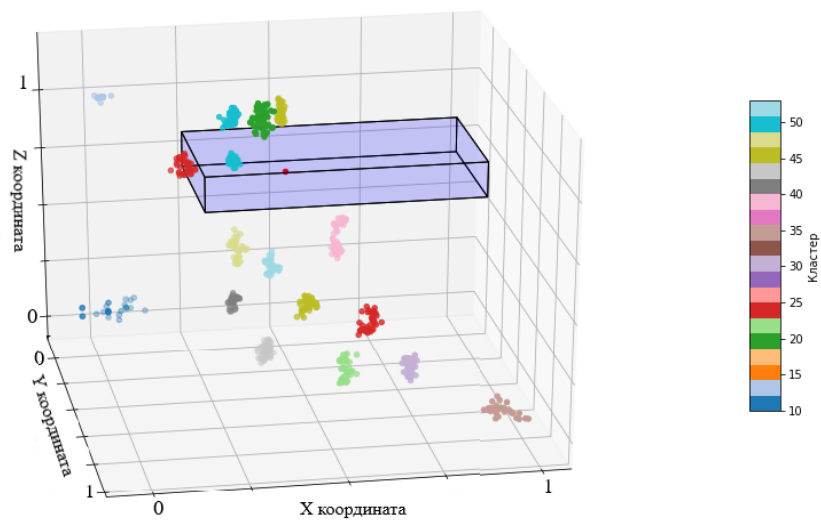

Рисунок 6 – Выделение всех «микрокластеров» (не более 100 событий в год) сейсмических событий. Фиолетовый паралеллепипед-опасная зона (зона горных работ+100 м по осям *Ox* и *Oy*, и 50 метров запаса по оси *Oz*) *(Составлен авторами)*

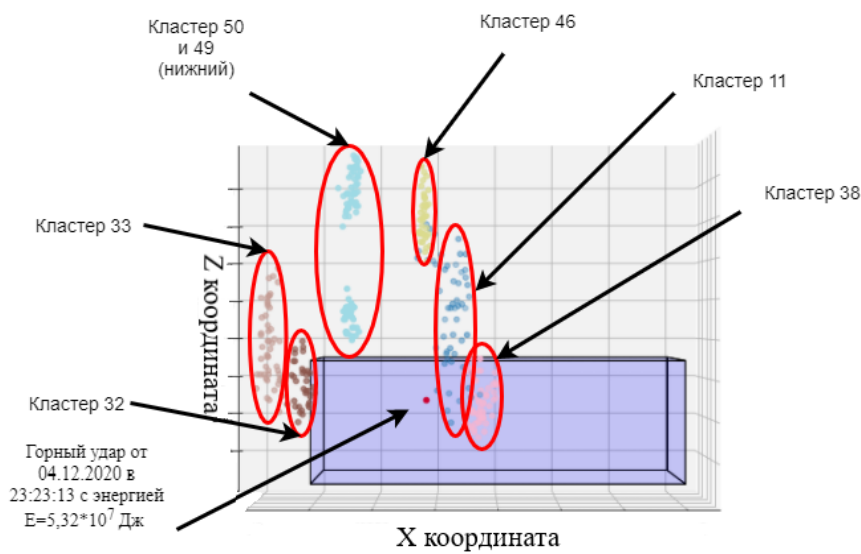

Рисунок 7 – Кластеры-"очаги разрушения горной породы" *(Составлен авторами)*

Для анализа результатов моделирования кластеров сейсмических событий С.А. Игнатьев [4] утверждает, что: «Кластерный анализ опирается на два основных предположения:
1) Выявленные признаки объекта должны допускать разбиение некоторой совокупности объектов на кластеры;
2) Правильность выбора должного масштаба или необходимых единиц величин, отражающих признаки объекта (в некоторых случаях требуется применение стандартизованных величин)».

В частности, авторы [32,33] предлагают использовать предварительную обработку исходных данных перед процедурой кластеризации. Исходные данные делятся на группы, каждая из которых характеризуется плотностью распределения сейсмических событий в пространстве сейсмического мониторинга. Данный подход позволит отфильтровать исходные данные от «шумовых» точек. Так же одним из этапов кластерного анализа является использование статистических критериев, полученных, в результате анализа данных сейсмического мониторинга Кукисвумчоррского месторождения за длительный период наблюдений (2000 – 2020 гг.), приведенных в табл. 3.

Таблица 3. **Статистические значимые критерии распределения сейсмических событий в ходе мониторинга Кукисвумчоррского месторождения за период 2000-2020 гг.**

| Название критерия | Значение в Джоулях |
|---|---|
| Среднее значение | 9620+/241.76 |
| Медиана | 1610 |
| Среднеквадратичное отклонение | 18000 |

*(Составлена авторами)*

Таким образом, все сейсмические события, превышающие выборочное среднее (на основании экспертной оценки сотрудников Научного центра геомеханики и проблем горного производства Санкт-Петербургского горного университета было выбрано значение 104 Дж), рассматриваются как сейсмические события – «предвестники», способные привести к опасным проявлениям ГД. Используя в качестве критериев качества предложенные в работе метрики, в ходе многокритериальной оптимизации [10,32,33] с помощью разработанного алгоритма были получены кластеры «неклассических» геометрических форм (эллипсоиды), которые условно можно поделить на 3 группы по расположению относительно произошедшего горного удара. Кластеры 32-33 находятся севернее точки удара, кластеры 50,49,46 находятся «сверху» над точкой удара. Кластеры 11 и 38 находятся на юго-востоке от точки удара. Следует отметить, что все выделенные кластеры являются «устойчивыми» во времени, то есть включают в себя сейсмические события в течении всего периода мониторинга в 2020 году. В табл. 4 приведены кластеры и содержащиеся в них события – «предвестники», которые с высокой долей вероятности могут являться триггерами горных ударов. Мониторинг сейсмических событий в этих кластерах в дальнейшем является целесообразным. В табл. 5 приведены результаты расчета критериев качества FAR/MAR, упомянутых в предыдущих разделах. В качестве событий для подсчета метрики используются события – «предвестники» из табл. 4. Обсуждение результатов анализа критериев приводится в заключении.

Таблица 4. **Характеристики сейсмических событий -«предвестников» выбранных кластеров, превышающих $10^4$ Дж**

| Время наступления сейсмического события | Величина энергии сейсмического события | Номер кластера | Время предполагаемого горного удара |
|---|---|---|---|
| 2020-03-11 22:13:27 | 976000 | 50 | 6 апреля 2020 |
| 2020-04-12 17:53:49 | 153000 | 50 | 21 апреля 2020 |
| 2020-05-21 02:55:17 | 406000 | 50 | 6 июня 2020 |
| 2020-06-07 21:27:14 | 94600 | 46 | 6 июня 2020 |
| 2020-06-15 06:33:30 | 18400 | 50 | 21 июня 2020 |
| 2020-07-05 18:20:15 | 362400 | 50 | 5 июля 2020 |
| 2020-11-08 08:31:49 | 22300 | 32 | 17 ноября 2020 |
| 2020-11-21 01:24:33 | 15000 | 32 | 3 декабря 2020 |
| 2020-11-21 02:23:31 | 27400 | 33 | 3 декабря 2020 |
| 2020-11-28 13:48:17 | 24000 | 32 | 3 декабря 2020 |

*(Составлена авторами)*

Таблица 5. **Характеристики сейсмических событий -«предвестников» выбранных кластеров, превышающих $10^4$ Дж**

| Время наступления сейсмического события | Значения критериев FAR/MAR. Краткосрочный горизонт прогноза | Значения критериев FAR/MAR. Среднесрочный горизонт прогноза | Значения критериев FAR/MAR. Долгосрочный горизонт прогноза |
|---|---|---|---|
| 08.04.2020 | 1/1 | 0/0 | 1/1 |
| 06.07.2020 | 3/1 | 3/0 | 3/1 |
| 04.12.2020 | 3/1 | 0/0 | 3/1 |

*(Составлена авторами)*

**Заключение**

Анализируя кластеризацию сейсмических событий на основании данных мониторинга за 2020 год, можно сделать следующие выводы:

1. Сформирована гипотетическая связь между опасным проявлением ГД и плавным снижением тренда сейсмической активности до точки глобального минимума с последующим изменением в направлении тренда. Для каждого из горных ударов были найдены свои события – «предвестники». Их количество, разница во времени и значения энергий тесно связаны с вероятностью наступления горного удара.

2. Предложена схема композитной модели анализа данных сейсмического мониторинга. На основании данных сейсмического мониторинга за 2020 год, были получены экспериментальные данные, подтверждающие сформированную авторами гипотезу. Полученные результаты являются развитием идеи, предложенными авторами в статье, об «использовании методов машинного обучения при работе с математическими моделями временных рядов, которые базируются на данных сейсмического мониторинга» [1].

3. Анализ полученных данных, их распределения во времени и пространстве, выявил даты двух потенциально возможных опасных проявлений ГД 06.04.2020 и 03.12.20, соответственно. При этом, фактические даты горных ударов от 08.04.2020 и 04.12.2020 отличаются от предполагаемых на 1-2 дня. В случае с реальным горным ударом от 06.07.2020 авторами был проведен дополнительный анализ временного ряда сейсмического мониторинга и внесены поправки в прогноз опасных проявлений ГД.

Проведем анализ критериев доли ложных (FAR) и пропущенных тревог (MAR):
- Краткосрочный горизонт прогнозирования: доля ложных срабатываний - 2,3 ложных тревоги на 1 горный удар, процент пропущенных горных ударов – 100 %;
- Среднесрочный горизонт прогнозирования: доля ложных срабатываний – 1 ложное событие на 1 горный удар, процент пропущенных горных ударов – 0 %;
- Долгосрочный горизонт прогнозирования: доля ложных срабатываний - 2,3 события на 1 горный удар, процент пропущенных горных ударов – 100 %;

Таким образом, предложенный метод зависит от выбора горизонта прогноза и является эффективным при использовании среднесрочного горизонта прогнозирования. Можно утверждать, что композитная модель, основанная на анализе временного ряда сейсмического мониторинга и распределении кластеров-"очагов" сейсмических событий в массиве горных пород, является эффективным средством контроля опасных проявлений ГД.